\documentclass{article}%
\usepackage{amsmath}
\usepackage{amsfonts}
\usepackage{amssymb}
\usepackage{graphicx}%
\providecommand{\U}[1]{\protect\rule{.1in}{.1in}}
\newtheorem{theorem}{Theorem} [section]

\newtheorem{conjecture}[theorem]{Conjecture}

\newtheorem{problem}[theorem]{Problem}
\newtheorem{proposition}[theorem]{Proposition}

\newenvironment{proof}[1][Proof]{\noindent\textbf{#1.} }{\ \rule{0.5em}{0.5em}}
\begin{document}

\author{Vadim E. Levit and David Tankus\\Department of Computer Science and Mathematics\\Ariel University Center of Samaria, ISRAEL\\\{levitv, davidta\}@ariel.ac.il}
\title{Weighted Well-Covered Graphs without $C_{4}$, $C_{5}$, $C_{6}$, $C_{7}$.}
\date{}
\maketitle

\begin{abstract}
A graph is \textit{well-covered} if every maximal independent set has the same
cardinality. The recognition problem of well-covered graphs is known to be
\textbf{co-NP}-complete. Let $w$ be a linear set function defined on the
vertices of $G$. Then $G$ is $w$\textit{-well-covered} if all maximal
independent sets of $G$ are of the same weight. The set of weight functions
$w$ for which a graph is $w$-well-covered is a \textit{vector space}. We prove
that finding the vector space of weight functions under which an input graph
is $w$-well-covered can be done in polynomial time, if the input graph
contains neither $C_{4}$ nor $C_{5}$ nor $C_{6}$ nor\ $C_{7}$.

\end{abstract}

\section{Introduction}

Throughout this paper $G=(V,E)$ is a simple (i.e., a finite, undirected,
loopless and without multiple edges) graph with vertex set $V=V(G)$ and edge
set $E=E(G).$

Cycles of $k$ vertices are denoted by $C_{k}$. When we say that $G$ does not
contain $C_{k}$ for some $k\geq3$, we mean that $G$ does not admit subgraphs
isomorphic to $C_{k}$. It is important to mention that these subgraphs are not
necessarily induced.

The graph $H$ is an \textit{induced} subgraph of $G$ if $V(H)\subseteq V(G)$,
and $E(H)$ coincides with set of all the edges that appear in $G$ over $V(H)$.
In this case $H=G[V(H)]$, and the graph $H$ is said to be\textit{ induced by
the set} $V(H)$.

Let $S\subseteq V$ be a set of vertices, and let $i\in\mathbb{N}$. Then
\[
N_{i}(S)=\{w\in V|\ \min_{s\in S}\ d(w,s)=i\},
\]
where $d(x,y)$ is the minimal number of edges required to construct a path
between $x$ and $y$. If $i\neq j$ then, obviously, $N_{i}(S)\cap
N_{j}(S)=\emptyset$. If $S=\left\{  v\right\}  $ for some $v\in V$, then
$N_{i}(\{v\})$ is simply denoted by $N_{i}(v)$.

A set of vertices $S\subseteq V$ is \textit{independent} if for every $x,y\in
S$, $x$ and $y$ are not adjacent. It is clear that an empty set is
independent. The \textit{independence number} of a graph $G$, denoted
$\alpha(G)$, is the size of a maximum cardinality independent set in $G$. A
graph is \textit{well-covered} if every maximal independent set has the same
cardinality, $\alpha(G)$. Finding the independence number of an input graph is
generally an \textbf{NP}-complete problem. However, if the input is restricted
to well-covered graphs then the problem can be solved polynomially by applying
the \textit{greedy algorithm}.

A well-covered graph $G$ is $1$\textit{-well-covered} if and only if for every
vertex $v\in G$, the graph $G-v$ is well-covered and $\alpha(G)=\alpha(G-v)$.

Let $T\subseteq V$. Then $S$ \textit{dominates} $T$ if $S\cup N_{1}%
(S)\supseteq T$. If $S$ and $T$ are both empty, then $N_{1}(S)=\emptyset$, and
therefore $S$ dominates $T$. If $S$ is a maximal independent set of $G$, then
it dominates $V(G)$.

Two adjacent vertices, $x$ and $y$, in $G$ are said to be \textit{related} if
there is an independent set $S$, containing neither $x$ nor $y$, such that
$S\cup\{x\}$ and $S\cup\{y\}$ are both maximal independent sets in the graph.
If $x$ and $y$ are related, then $xy$ is a \textit{relating edge}. It is
proved in \cite{bnz:related} that deciding whether an edge in an input graph
is relating is an \textbf{NP}-complete problem.

\begin{theorem}
\cite{bnz:related} \label{relatednpc} The following problem is \textbf{NP}-complete:

Input: A graph $G=(V,E)$ and an edge $xy\in E$.

Question: Is $xy$ a relating edge?
\end{theorem}

However, if the input graph contains neither $C_{4}$ nor $C_{6}$ then the
problem is polynomial.

\begin{theorem}
\cite{lt:wcc46} \label{related46} The following problem is polynomially solvable:

Input: A graph $G=(V,E)$, which contains neither $C_{4}$ nor $C_{6}$, and an
edge $xy\in E$.

Question: Is $xy$ a relating edge?
\end{theorem}

The recognition of well-covered graphs is known to be \textbf{co-NP}-complete
\cite{chsl1993}, \cite{sast1992}. The problem remains \textbf{co-NP}-complete
even when the input graph is $K_{1,4}$-free \cite{cst:structures}. However,
the problem is polynomially solvable for $K_{1,3}$-free graphs
\cite{tata:wck13f,tata:wck13fn}, for graphs with girth at least $5$
\cite{fhn:wcg5}, for graphs that contain neither $C_{4}$ nor $C_{5}$
\cite{fhn:wc45}, for graphs with a bounded maximum degree \cite{cer:degree},
for perfect graphs of bounded clique size \cite{deanzito1994}, or for chordal
graphs \cite{ptv:chordal}. Recognizing $1$-well-covered graphs with no $C_{4}$
can be implemented in polynomial time as well \cite{h:1wc4}.

Brown, Nowakowski and Zverovich investigated well-covered graphs with no
$C_{4}$, and presented the following open problem.

\begin{problem}
\cite{bnz:related} \label{wcc4} What is the complexity of determining whether
an input graph with no $C_{4}$ is well-covered?
\end{problem}

Levit and Tankus proposed the following.

\begin{conjecture}
\cite{lt:wcc46} \label{wcc46} The following problem can be solved in
polynomial time:

Input: A graph $G=(V,E)$ which contains neither $C_{4}$ nor $C_{6}$.

Question: Is $G$ well-covered?
\end{conjecture}

The main finding of this article is a polynomial time algorithm, which
receives as its input a graph without $C_{4}$, $C_{5}$, $C_{6}$ and $C_{7}$,
and finds the vector space of weight functions $w$ for which the graph is $w$-well-covered.

In the Section 2 we define the notion of \textit{generating subgraphs}. Then
we prove that given an input graph $G$ which contains neither $C_{4}$ nor
$C_{6}$ nor $C_{7}$, for each induced complete bipartite subgraph $B$ of $G$
it can be decided polynomially whether $B$ is generating.

In Section 3 we consider the fact that a weighted graph $(G;w)$ with the
family of all its independent sets forms a \textit{weighted hereditary
system}. This system is greedy if and only if $G$ is $w$-well-covered. We
quote several known results about greedy hereditary systems, and use them to
prove that a graph is $w$-well-covered if and only if it satisfies all the
constraints produced by its generating subgraphs.

In Section 4 we consider the fact that the set of all weight functions $w$ for
which a graph is $w$-well-covered is a \textit{vector space}. We prove that
finding that vector space can be done polynomially if the input is restricted
to graphs without $C_{4}$, $C_{5}$, $C_{6}$ and $C_{7}$.

In the Conclusions we present an open problem for further research.

\section{Generating Subgraphs}

\label{generating} In this section we define the notion of a
\textit{generating subgraph}. Then we prove that given an input graph $G$
without $C_{4}$, $C_{6}$ and $C_{7}$, and an induced complete bipartite
subgraph $B$ of $G$, it can be decided polynomially whether $B$ is generating.

Let $G=(V,E)$ be a graph, and let $w:V\longrightarrow\mathbb{R}$ be a weight
function defined on its vertices. The weight of a set $S\subseteq V$ is
defined by: $w(S)=\sum_{s\in S}w(s)$. The graph $G$ is $w$%
\textit{-well-covered} if all maximal independent sets of $G$ are of the same
weight \cite{cer:degree}.

Let $B$ be a complete bipartite induced subgraph of $G$, and denote the vertex
sets of the bipartition of $B$ by $B_{X}$ and $B_{Y}$. Then $B$ is a
\textit{generating} subgraph of $G$ if there exists an independent set $S$ of
$G$ such that $S\cup B_{X}$ and $S\cup B_{Y}$ are both maximal independent
sets of $G$. In this case $B$ \textit{produces} the constraint that $B_{X}$
and $B_{Y}$ are of the same weight. If $B$ is a generating subgraph of $G$,
and $w$ is a weight function defined on the vertices of $G$ such that
$w(B_{X})=w(B_{Y})$ then $w$ \textit{satisfies the constraint produced by }$B$.

When a subgraph $B$ is isomorphic to $K_{1,1}$, it is generating if and only
if its two vertices are related. Hence the notion of related vertices is an
instance of a generating subgraph.

For every $P\in\{B_{X},B_{Y}\}$, let $Q=V(B)\diagdown P$, and define
\[
M_{1}(P)=N_{1}(P)\cap N_{2}(Q),M_{2}(P)=N_{1}(M_{1}(P))\diagdown B.
\]

\begin{proposition}
\label{prop1}The subgraph $B$ is generating if and only if there exists an
independent subset of the set $N_{2}(B)$, which dominates $N_{1}\left(
B_{X}\right)  \bigtriangleup N_{1}\left(  B_{Y}\right)  =M_{1}(B_{X})\cup
M_{1}(B_{Y})$.
\end{proposition}

\begin{proof}
Assume that there exists an independent subset $S$ of the set $N_{2}(V(B))$,
which dominates $M_{1}(B_{X})\cup M_{1}(B_{Y})$. Let us expand it arbitrarily
to a maximal independent set $S^{\ast}$ in $V\left(  G\right)  \diagdown
\left(  V\left(  B\right)  \cup N_{1}(V(B))\right)  $. Then $S^{\ast}\cup
B_{X}$ and $S^{\ast}\cup B_{Y}$ are maximal independent sets of $G$. Thus, by
definition, $V(B)=\left(  S^{\ast}\cup B_{X}\right)  \bigtriangleup\left(
S^{\ast}\cup B_{Y}\right)  $ is generating.

Conversely, assume that the subgraph $B$ is generating. Then there exist two
maximal independent sets, $S_{1}$ and $S_{2}$ of $G$ such that $S_{1}%
\bigtriangleup S_{2}=V(B)$. Therefore, $S_{1}\cap S_{2}\cap(N_{2}(V(B)))$ is
an independent set of $G$\ that dominates $M_{1}(B_{X})\cup M_{1}(B_{Y})$.
\end{proof}

If $B$ is generating, then every independent set $S\subseteq V\left(
G\right)  \diagdown(V\left(  B\right)  \cup N_{1}(V(B)))$ that dominates
$N_{1}(B_{X})\bigtriangleup N_{1}(B_{Y})$ is called a \textit{witness} of the
fact that $B$ is generating. According to Proposition \ref{prop1}, for every
generating subgraph $B$, there exists a witness $S\subseteq N_{2}(V(B))$.

The following theorem is a generalization of Theorem \ref{related46} for the
case that the input graph does not contain $C_{7}$.

\begin{theorem}
\label{generating1} The following problem can be solved in polynomial time:

Input: A graph $G=(V,E)$ which contains neither $C_{4}$ nor $C_{6}$ nor
$C_{7}$, and a complete bipartite induced subgraph $B$ of $G$.

Question: Is $B$ a generating subgraph of $G$?
\end{theorem}

\begin{proof}
Let us recall that the vertex sets of the bipartition of $B$ are denoted by
$B_{X}$ and $B_{Y}$. Assume, without loss of generality, that $\left\vert
B_{X}\right\vert \leq\left\vert B_{Y}\right\vert $. Notice that since the
graph $G$ does not contain $C_{4}$, the set $B_{X}$ contains just one element,
i.e., $\left\vert B_{X}\right\vert =1$.

Let $B_{X}=\{x\}$ and $B_{Y}=\{y_{1},...,y_{k}\}$. Since $G$ contains neither
$C_{4}$ nor $C_{6}$ nor $C_{7}$, we obtain the following:

\begin{itemize}
\item $\forall1 \leq i < j \leq k \ \ N_{1}(y_{i}) \cap N_{1}(y_{j}) = \{ x
\}$. (If $N_{1}(y_{i}) \cap N_{1}(y_{j})$ contains another vertex, $v$, then
$(x,y_{i},v,y_{j})$ is a $C_{4}$.)

\item $\forall1 \leq i < j \leq k \ \ N_{2}(y_{i}) \cap N_{2}(y_{j}) \cap
N_{3}(x) = \emptyset$. (If there exists $v \in N_{2}(y_{i}) \cap N_{2}(y_{j})
\cap N_{3}(x)$, then there are two edge disjoint 2-length paths from $y_{i}$
and $y_{j}$ to $v$. The vertices of these paths and the vertex $x$ are on a
$C_{6}$.)

\item For every $1 \leq i < j \leq k$ there are no edges between $N_{2}(y_{i})
\cap N_{3}(x)$ and $N_{2}(y_{j}) \cap N_{3}(x)$. (Assume that $v_{i} \in
N_{2}(y_{i}) \cap N_{3}(x)$ and $v_{j} \in N_{2}(y_{j}) \cap N_{3}(x)$ are
adjacent. Then $y_{i}$, $v_{i}$, $v_{j}$, $y_{i}$ and $x$ are on a $C_{7}$.)

\item For every $1 \leq i \leq k$, every connected component of $N_{1}(y_{i})
\cap N_{2}(x)$ contains at most one edge. (If a connected component of
$N_{1}(y_{i}) \cap N_{2}(x)$ contains a path $(v_{1}, v_{2}, v_{3})$, then
$(v_{1}, v_{2}, v_{3}, y_{i})$ is a $C_{4}$.)

\item Every connected component of $N_{3}(x)$ contains at most one edge.
(Assume, on the contrary, that a connected component of $N_{3}(x)$ contains a
path $(v_{1},v_{2},v_{3})$. Let $P_{1}$ and $P_{3}$ be shortest paths from
$v_{1}$ and $v_{3}$ to $x$, respectively. Let $v$ be the first vertex in the
intersection of $P_{1}$ and $P_{3}$. The vertices $v_{1}$, $v_{2}$, $v_{3}$
and $x$ are on a forbidden cycle: If $v=x$ they are on a $C_{7}$. If $v\in
N_{1}(x)$, they are on a $C_{6}$. In the remaining case $v\in N_{2}(x)$, and
they are on a $C_{4}$.)

\item Every vertex of $N_{3}(x)$ is adjacent to exactly one vertex of
$N_{2}(x)$. (Assume, on the contrary, that a vertex $v\in N_{3}(x)$ is
adjacent to two distinct vertices, $v_{1}$ and $v_{2}$, of $N_{2}(x)$. If
$N_{1}(v_{1})\cap N_{1}(v_{2})\cap N_{1}(x)=\emptyset$ then $v_{1}$, $v$,
$v_{2}$ and $x$ are on a $C_{6}$. Otherwise $v_{1}$, $v$ and $v_{2}$ are on a
$C_{4}$.)
\end{itemize}

The fact that the graph $G$ does not contain $C_{6}$ implies the following:

\begin{itemize}
\item There are no edges connecting vertices of $M_{2}(B_{X})$ with vertices
of $M_{2}(B_{Y})$.

\item The set $M_{2}(B_{X}) \cap M_{2}(B_{Y})$ is independent.

\item There are no edges between the vertices belonging to $M_{2}(B_{X})\cap
M_{2}(B_{Y})$ and the other vertices of $M_{2}(B_{X})\cup M_{2}(B_{Y})$.
\end{itemize}

Consequently, if $S_{x}\subseteq M_{2}(B_{X})$ and $S_{y}\subseteq M_{2}%
(B_{Y})$ are independent, then $S_{x}\cup S_{y}$ is independent as well.
Therefore, by Proposition \ref{prop1} it is enough to prove that one can
decide in polynomial time whether there exists an independent subset of the
set $M_{2}(P)$ dominating $M_{1}(P)$, where $P\in\{B_{X},B_{Y}\}$.

Let us note that:

\begin{itemize}
\item Every vertex of $M_{2}(P)$ is adjacent to exactly one vertex of
$M_{1}(P)$, or otherwise the graph $G$ contains a $C_{4}$.

\item Every connected component of $M_{2}(P)$ contains at most $2$ vertices,
or otherwise the graph $G$ contains either $C_{4}$ or $C_{6}$ or $C_{7}$.
\end{itemize}

Let $A_{1},...,A_{k}$ be the connected components of $M_{2}(P)$. Define a flow
network
\[
F_{P}=\{G_{F}=(V_{F},E_{F}),s\in V_{F},t\in V_{F},w:E_{F}\longrightarrow
\mathbb{R}\}
\]
as follows.

Let
\[
V_{F}=M_{1}(P)\cup M_{2}(P)\cup\{a_{1},...,a_{k},s,t\},
\]
where $a_{1},...,a_{k},s,t$ are new vertices, $s$ and $t$ are the source and
sink of the network, respectively.

The directed edges $E_{F}$ are:

\begin{itemize}
\item the directed edges from $s$ to each vertex of $M_{1}(P)$;

\item all directed edges $v_{1}v_{2}$ s.t. $v_{1}\in M_{1}(P)$, $v_{2}\in
M_{2}(P)$ and $v_{1}v_{2}\in E$;

\item the directed edges $va_{i}$, for each $1\leq i\leq k$ and for each $v\in
A_{i}$;

\item the directed edges $a_{i}t$, for each $1\leq i\leq k$.
\end{itemize}

Let $w\equiv1$. Invoke any polynomial time algorithm for finding a maximum
flow in the network, for example, Ford and Fulkerson's algorithm. Let $S_{P}$
be the set of vertices in $M_{2}(P)$ in which there is a positive flow.

Assume, on the contrary, that $S_{P}$ is not independent. There exist two
adjacent vertices, $v_{1}$ and $v_{2}$, in $S_{P}$. Hence $v_{1}$ and $v_{2}$
belong to the same connected component $A_{i}$ of $M_{2}(P)$. Therefore there
exist a flow of size 1 on each of the directed edges $v_{1}a_{i}$ and
$v_{2}a_{i}$ in the network. There exist a flow of size at least 2 in the edge
$a_{i}t$, which is a contradiction to the fact that all edges in the network
have capacity 1. Therefore $S_{P}$ is independent.

The maximality of $S_{P}$ implies that $\left\vert M_{1}(P)\cap N_{1}%
(S_{P})\right\vert \geq|M_{1}(P)\cap N_{1}(S_{P}^{\prime})|$, for any
independent set $S_{P}^{\prime}$ of $M_{2}(P)$.

Let us conclude the proof with the recognition algorithm for generating subgraphs.

For each $P\in\{B_{X},B_{Y}\}$, build a flow network $F_{P}$, and find a
maximum flow. Let $S_{P}$ be the set of vertices in $M_{2}(P)$ in which there
is a positive flow. If $S_{P}$ does not dominate $M_{1}(P)$ the algorithm
terminates announcing that $B$ is not generating. Otherwise, let $S$ be any
maximal independent set in $V\left(  G\right)  \diagdown V\left(  B\right)  $
which contains $S_{B_{X}}\cup S_{B_{Y}}$. Each of $S\cup B_{X}$ and $S\cup
B_{Y}$ is a maximal independent set of $G$, and $B$ is generating.

This algorithm can be implemented in polynomial time: One iteration of Ford
and Fulkerson's algorithm includes:

\begin{itemize}
\item Updating the flow function. (In the first iteration the flow equals $0$.)

\item Constructing the residual graph.

\item Finding an augmenting path, if exists. The residual capacity of every
augmenting path is equal to $1$.
\end{itemize}

Each of the above can be implemented in $O\left(  \left\vert V\right\vert
+\left\vert E\right\vert \right)  $ time. In each iteration the number of
vertices in $M_{2}(P)$ with a positive flow increases by 1. Therefore, the
number of iterations can not exceed $|V|$, and Ford and Fulkerson's algorithm
terminates in $O\left(  \left\vert V\right\vert \left(  \left\vert
V\right\vert +\left\vert E\right\vert \right)  \right)  $ time. Our algorithm
invokes Ford and Fulkerson's algorithm twice, and terminates in $O\left(
\left\vert V\right\vert \left(  \left\vert V\right\vert +\left\vert
E\right\vert \right)  \right)  $ time.
\end{proof}

\section{Hereditary Systems}

\label{hereditary} In this Section we introduce the notion of a
\textit{hereditary system}. We quote several known results about greedy
hereditary systems, and use them to prove that a weighted graph $(G;w)$ is
$w$-well-covered if and only if it satisfies all the constraints produced by
its generating subgraphs.

A \textit{hereditary system} is a pair $H=(S,F)$, where $S$ is a finite set
and $F$ is a family of subsets of $S$, where $f\in F$ and $f^{\prime}\subseteq
f$ implies $f^{\prime}\in F$. The members of $F$ are called \textit{feasible}
sets of the system.

A \textit{weighted hereditary system} is a pair $(H,w)$, where $H=(S,F)$ is a
hereditary system, and $w:S\longrightarrow\mathbb{R}$ is a weight function on
$S$. The weight of a set $S^{\prime}\subseteq S$ is defined by:
\[
w(S^{\prime})=%
{\displaystyle\sum\limits_{s^{\prime}\in S^{\prime}}}
w(s^{\prime}).
\]

A \textit{greedy weighted hereditary system} is a weighted hereditary system
$(H,w)$ for which all maximal feasible sets are of the same weight.

\begin{theorem}
\cite{tata:hs} \label{hsdelta} Let
\[
(H=\left(  S,F\right)  ,w:S\longrightarrow\mathbb{R})
\]
be a weighted hereditary system. Then $(H,w)$ is not greedy if and only if
there exist two maximal feasible sets, $S_{1}$ and $S_{2}$, of $F$ with
different weights, $w(S_{1})\neq w(S_{2})$, such that for each $a\in
S_{1}\setminus S_{2}$ and for each $b\in S_{2}\setminus$ $S_{1}$, the set
$(S_{1}\cap S_{2})\cup\{a,b\}$ is not feasible.
\end{theorem}

Let $\left(  G,w\right)  =\left(  V,E,w:V\longrightarrow\mathbb{R}\right)  $
be a weighted graph. Then $(G,w)$ with the family of all its independent sets
clearly forms a weighted hereditary system. This system is greedy if and only
if $G$ is $w$-well-covered. Hence, the following is an instance of Theorem
\ref{hsdelta}.

\begin{theorem}
\cite{tata:hs} \label{isdelta} Let $(G,w)$ be a weighted graph. Then $G$ is
not $w$-well-covered if and only if there exist two maximal independent sets,
$S_{1}$ and $S_{2}$ of $G$ with different weights $w(S_{1})\neq w(S_{2})$,
such that $G\left[  S_{1}\bigtriangleup S_{2}\right]  $ is complete bipartite.
\end{theorem}

We now state and prove the following.

\begin{theorem}
\label{space} Let $\left(  G,w\right)  $ be a weighted graph. Then $G$ is
$w$-well-covered if and only if it satisfies all the constraints produced by
generating subgraphs of $G$.
\end{theorem}

\begin{proof}
According to the definition of a generating subgraph, if $G$ is $w$%
-well-covered and $B$ is a generating subgraph of $G$, then the vertex sets of
the bipartition of $B$ must have equal weights.

Assume that $G$ is not $w$-well-covered. By Theorem \ref{isdelta}, there exist
two maximal independent sets $S_{1}$ and $S_{2}$ of $G$ such that
$w(S_{1})\neq w(S_{2})$, and the subgraph $G\left[  S_{1}\bigtriangleup
S_{2}\right]  $ is complete bipartite. Let $H=G\left[  S_{1}\bigtriangleup
S_{2}\right]  $ be a complete bipartite subgraph. The union of $S_{1}\cap
S_{2}$ with either vertex set of the bipartition of $H$ is a maximal
independent set of the graph. Therefore, $H$ is generating.
\end{proof}

\section{The Vector Space}

The set of all weight functions $w:V\longrightarrow\mathbb{R}$ for which the
graph $G=(V,E)$ is $w$-well-covered is a \textit{vector space}
\cite{cer:degree}. Assume that $G$ contains neither $C_{4}$ nor $C_{5}$ nor
$C_{6}$ nor $C_{7}$. In Section \ref{generating} we proved that for every
complete bipartite induced subgraph $B$ of $G$ it is possible to decide in
polynomial time whether $B$ is generating. In Section \ref{hereditary} it was
shown that the union of constraints produced by all generating subgraphs of
$G$ is the vector space of weight functions under which $G$ is $w$%
-well-covered. However, the number of generating subgraphs of $G$ is not
necessarily polynomial. In this section we supply an algorithm to find the
requested vector space in polynomial time.

For every $v\in V$, define $L_{v}$ to be the vector space of weight functions
of $G$ satisfying the union of all constraints produced by subgraphs $B$ of
$G$ with $B_{X}=\{v\}$. Suppose that $w$ is a weight function defined on $V$.
Then $G$ is $w$-well-covered if and only if $w\in%
{\displaystyle\bigcap\limits_{v\in V}}
L_{v}$.

\begin{theorem}
\label{generatingfamily} Let $G=(V,E)$ be a graph that contains neither
$C_{4}$ nor $C_{5}$ nor $C_{6}$ nor $C_{7}$. For every $v\in V$ it is possible
to find $L_{v}$ in polynomial time.
\end{theorem}

\begin{proof}
Let $v\in V$. For every non-empty vertex set $S \subseteq N_{1}(v)$, define
\[
M_{1}(S)=N_{1}(S) \cap N_{2}(v) \text{ and }M_{2}(S)=N_{2}(S)\cap N_{3}(v).
\]
If $S = \{y\}$ is a single vertex, then $M_{1}(\{y\})$ and $M_{2}(\{y\})$ will
be abbreviated to $M_{1}(y)$ and $M_{2}(y)$, respectively. Let $D(v)$ be the
set of all vertices $y$ of $N_{1}(v)$ such that there exists an independent
set of $M_{2}(y)$ which dominates $M_{1}(y)$. Note that $y\in D(v)$ if and
only if $v$ and $y$ are related in the subgraph of $G$ induced by $\{v,y\}\cup
M_{1}(y)\cup M_{2}(y)$. Hence it is possible to find $D(v)$ by invoking the
algorithm presented in the proof of Theorem \ref{generating1} for each $y\in
N_{1}(v)$.

Our first step is to build a family $\left\{  F_{v}\right\}  _{v\in V}$ of
generating subgraphs such that all the constraints produced by the members of
$F_{v}$ span the vector space $L_{v}$.

For every $v\in V$, define $F_{v}$ to be the family of the following bipartite
subgraphs of $G$:

\begin{itemize}
\item $B^{\ast}\in F_{v}$, where $B^{\ast}$ is a graph with the following
bipartition: $B_{X}^{\ast}=\{v\}$ and $B_{Y}^{\ast}$ containing exactly one
vertex from every connected component of $D(v)$.

\item If $B^{\ast}$ is not a copy of $K_{1,1}$, then $B^{\ast}\diagdown
\{y\}\in F_{v}$ for every $y\in B_{Y}^{\ast}\cap N_{1}(N_{2}(v))$.

\item $G\left[  V\left(  B^{\ast}\right)  \bigtriangleup V\left(  C\right)
\right]  \in F_{v}$ for every connected component $C$ of $D(v)$ such that
$\left\vert V\left(  C\right)  \right\vert =2$.
\end{itemize}

Since $G$ does not contain $C_{4}$, every connected component of $D(v)$
contains $2$ vertices at most. For every $y\in D(v)$, let $S_{y}$ be an
independent set of $M_{2}(y)$, which dominates $M_{1}(y)$. Let us prove that
$\bigcup_{y\in D(v)}S_{y}$ is independent. Suppose, on the contrary, that
there exist two adjacent vertices $a_{1},a_{2}\in$ $\bigcup_{y\in D(v)}S_{y}$.
It means that there are $y_{1},y_{2}\in D(v)$, $y_{1}\neq y_{2}$ such that
$a_{1}\in S_{y_{1}}\subseteq M_{2}(y_{1})$ and $a_{2}\in S_{y_{2}}\subseteq
M_{2}(y_{2})$. Since $d(y_{1},a_{1})=$ $d(y_{2},a_{2})=2$, the vertices
$v,y_{1},a_{1},a_{2},y_{2}$ belong to $C_{7}$, which contradicts the
hypothesis of the theorem.

For every vertex $y\in N_{1}(v)$ we define a set $m(y)$ of size at most $1$:
If $M_{1}(y)\neq\emptyset$ then $m(y)$ is a single vertex of $M_{1}(y)$,
otherwise $m(y)=\emptyset$. If $y_{1},y_{2}\in N_{1}(v)$, then $m(y_{1})\cup
m(y_{2})$ is independent, because $G$ does not contain $C_{5}$, and $S_{y_{1}%
}\cup m(y_{2})$ is an independent set, since $G$ does not contain $C_{6}$.

For every member of $F_{v}$ we present a witness that it is generating:

\begin{itemize}
\item A witness of $B^{\ast}$ is:
\[
S^{\ast}=(\bigcup_{y\in B_{Y}^{\ast}}S_{y})\cup(\bigcup_{y\in(N_{1}%
(v)\diagdown B_{Y}^{\ast})}m(y)).
\]

\item Let $y\in B_{Y}^{\ast}\cap N_{1}(N_{2}(v))$. A witness of $B^{\ast
}\diagdown\{y\}$ is:
\[
(S^{\ast}\diagdown S_{y})\cup m(y).
\]

\item Let $C=\{y_{1},y_{2}\}$ be a connected component of $D(v)$ with $2$
vertices, and assume, without loss of generality, that $y_{1}\in B_{Y}^{\ast}%
$. A witness of $G\left[  V\left(  B^{\ast}\right)  \bigtriangleup V\left(
C\right)  \right]  $ is:
\[
(S^{\ast}\diagdown(S_{y_{1}}\cup m(y_{2})))\cup S_{y_{2}}\cup m(y_{1}).
\]

\end{itemize}

Our next step is to prove that the constraint produced by every generating
subgraph $B$ of $G$ with $B_{X}=\{v\}$ is dependent on the constraints
produced by the members of $F_{v}$. Indeed, let $B$ be a generating subgraph
of $G$ with $B_{X}=\{v\}$. Then there exist connected components
$C_{1},...,C_{p}$ of $D(v)$ of size $2$, and members $y_{1},...,y_{q}$ of
$D(v)$, such that
\[
B=G\left[  (V\left(  B^{\ast}\right)  \bigtriangleup(\bigcup_{1\leq j\leq
p}V\left(  C_{j}\right)  ))\diagdown\{y_{i}|1\leq i\leq q\}\right]  .
\]
Then the constraint produced by $B$ is dependent on the constraints of the
following members of $F_{v}$:

\begin{itemize}
\item $B^{\ast}$.

\item $B^{\ast}\diagdown\{y_{i}\}$ for every $1 \leq i \leq q.$

\item $G\left[  V\left(  B^{\ast}\right)  \bigtriangleup V\left(
C_{j}\right)  \right]  $ for every $1\leq j\leq p$.
\end{itemize}

A weight function $w$ belongs to the vector space $L_{v}$ if and only if it
obeys the following constraints:

\begin{itemize}
\item $w(B_{X}^{\ast})=w(B_{Y}^{\ast})$. (The constraint produced by $B^{\ast
}$.)

\item If $B^{\ast}$ is not a copy of $K_{1,1}$ then $w(y)=0$ for every vertex
$y$ which is a connected component of $D(v)$ of size $1$. (A linear
combination of the constraints produced by $B^{\ast}$ and $B^{\ast}%
\diagdown\{y\}$.)

\item If $B^{\ast}$ is not a copy of $K_{1,1}$ then $w(y_{1})=w(y_{2})$ for
every connected component $C_{j}=\{y_{1},y_{2}\}$ of $D(v)$ of size $2$. (A
linear combination of the constraints produced by $B^{\ast}$ and $G\left[
V\left(  B^{\ast}\right)  \bigtriangleup V\left(  C_{j}\right)  \right]  $.)
\end{itemize}

The algorithm of finding a spanning set of $L_{v}$ is completed.

For each $y\in N_{1}(v)$ the decision whether $y\in D(v)$ takes $O\left(
\left\vert V\right\vert \left(  \left\vert V\right\vert +\left\vert
E\right\vert \right)  \right)  $ time. Hence, $D(v)$ can be found in $O\left(
\left\vert V\right\vert ^{2}\left(  \left\vert V\right\vert +\left\vert
E\right\vert \right)  \right)  $ time. In summary, the complexity of the
algorithm finding $L_{v}$ is $O\left(  \left\vert V\right\vert ^{2}\left(
\left\vert V\right\vert +\left\vert E\right\vert \right)  \right)  $.
\end{proof}

\begin{theorem}
\label{wwc} Let $G=(V,E)$ be a graph that contains neither $C_{4}$ nor $C_{5}$
nor $C_{6}$ nor $C_{7}$. Then it is possible to find in polynomial time the
vector space of weight functions $w$ under which the graph $G$ is $w$-well-covered.
\end{theorem}

\begin{proof}
According to Theorem \ref{space}, the vector space of weight functions of $G$
under which the graph is $w$-well-covered is the maximum linear subspace
satisfying all the constraints produced by generating subgraphs of $G$. Since
$G$ does not contain $C_{4}$, one of the vertex sets of the bipartition of
every generating subgraph comprises only one vertex. Hence, the required
vector space is $\bigcap_{v\in V}L_{v}$.

By Theorem \ref{generatingfamily}, for every $v\in V$ one can find $L_{v}$ in
$O\left(  \left\vert V\right\vert ^{2}\left(  \left\vert V\right\vert
+\left\vert E\right\vert \right)  \right)  $ time. Consequently, $\left\{
L_{v_{i}}|1\leq i\leq\left\vert V\right\vert \right\}  $ can be found in
$O\left(  \left\vert V\right\vert ^{3}\left(  \left\vert V\right\vert
+\left\vert E\right\vert \right)  \right)  $ time. In order to find the
intersection $\bigcap_{v\in V}L_{v}$, which is the vector space of weight
functions under which the graph is $w$-well-covered, it is enough to apply the
Gaussian elimination procedure to a matrix of size $\left(  \sum
_{i=1}^{\left\vert V\right\vert }g_{i}\right)  \bullet\left\vert V\right\vert
$, where $g_{i}$ is the number of generating subgraphs of $G$ belonging to
$F_{v_{i}}$. Since $\sum_{i=1}^{\left\vert V\right\vert }g_{i}\leq\left\vert
V\right\vert ^{2}$, the time complexity of the Gaussian elimination procedure
for this matrix is bounded by $O\left(  \left\vert V\right\vert ^{4}\right)
$. Finally, $\bigcap_{v\in V}L_{v}$ may be constructed in $O\left(  \left\vert
V\right\vert ^{3}\left(  \left\vert V\right\vert +\left\vert E\right\vert
\right)  \right)  $.
\end{proof}

\section{Conclusions}

An important question for any family of graphs is: Does there exist an
efficient recognition algorithm? The fact that well-covered graph recognition
is co-\textbf{NP}-complete was proved independently in \cite{chsl1993} and
\cite{sast1992}. A more challenging problem is how to find the vector space of
weight functions allowing a graph to be $w$-well-covered.

Our main conjecture reads as follows.

\begin{conjecture}
\label{conj1}The following problem can be solved in polynomial time:

Input: A graph $G$ which contains neither $C_{4}$ nor $C_{6}$ nor $C_{7}$.

Question: Find the vector space of weight functions $w$ under which the graph
$G$ is $w$-well-covered.
\end{conjecture}

\section{Acknowledgment}

We thank the two anonymous referees for their valuable comments that helped us
to improve the presentation of the paper. Our special thanks to one referee
for a question that led us to revealing an error in the first version of the
proof of Theorem \ref{generatingfamily}. While Theorem \ref{generatingfamily}
in the fixed from is presented in this paper, the first version turns out to
be Conjecture \ref{conj1}.

\end{document}